# Transient Heavy Element Absorption Systems in Novae: Episodic Mass Ejection from the Secondary Star

Robert Williams[1], Elena Mason[2], Massimo Della Valle[3], & Alessandro Ederoclite[2,4]

## ABSTRACT

A high-resolution spectroscopic survey of postoutburst novae reveals short-lived heavy element absorption systems in a majority of novae near maximum light, having expansion velocities of 400–1000 km s$^{-1}$ and velocity dispersions between 35-350 km s$^{-1}$. A majority of systems are accelerated outwardly, and they all progressively weaken and disappear over timescales of weeks. A few of the systems having narrow, deeper absorption reveal a rich spectrum of singly ionized Sc, Ti, V, Cr, Fe, Sr, Y, Zr, and Ba lines. Analysis of the richest such system, in Nova LMC 2005, shows the excitation temperature to be $10^4$ K and elements lighter than Fe to have abundance enhancements over solar values by up to an order of magnitude. The gas causing the absorption systems must be circumbinary and its origin is most likely mass ejection from the secondary star. The absorbing gas pre-exists the outburst and may represent episodic mass transfer events from the secondary star that initiate the nova outburst(s). If SNe Ia originate in single degenerate binaries, such absorption systems could be detectable before maximum light.

*Subject headings*: Novae, cataclysmic variables

*Short Title:* Novae Transient Absorption Systems

---

[1] Space Telescope Science Institute, 3700 San Martin Drive, Baltimore, MD 21218; wms@stsci.edu
[2] European Southern Observatory, Alonso de Cordova 3107, Vitacura, Santiago, Chile
[3] Osservatorio Astronomico di Arcetri, Largo E. Fermi 5, Firenze, Italy
[4] Current address: Instituto de Astrofisica de Canarias, Vía Láctea s/n, E-38205 La Laguna, Tenerife, Spain

## 1. Introduction

We have obtained sequences of high-resolution (R=48,000) optical spectra of a number of novae in the months following their outbursts with the FEROS echelle spectrograph on the ESO La Silla 1.5m and 2.2m telescopes. These spectra are a rich source of information about the outburst and ejecta. Their primary limitation lies in the fact that because of fiber transmissivity they do not extend below 3900 Å, where the primary signatures of He/N or 'neon' novae occur, and few of our spectra were obtained in the fainter decline stage more than 6-8 months after the outburst because of the moderate apertures of the telescopes, thus failing to sample the later nebular stages of the ejecta.

The spectral evolution of several of the novae in our survey, viz., V382 Vel/99, SMC 2001, LMC 2002, and V5114 Sgr/04 have already been discussed in detail in earlier papers in this series (Della Valle et al. 2002; Mason et al. 2005; Ederoclite et al. 2006). FEROS is a bench-mounted fiber-fed instrument with a fixed format (Kaufer et al. 1999), so except for different intensity signal-to-noise ratios that vary because of differing object brightness and exposure times, the quality of the novae spectra displayed in the above papers is typical of the data for all of our objects. We are currently analyzing these data, and we discuss here initial results of the analysis of absorption lines that we have found to appear in the early decline phase of the majority of the novae we have observed.

## 2. The Spectroscopic Data

Our present targets consist of Galactic and Magellanic Cloud novae that were discovered in the interval late 2003 to early 2006 that were observable from the southern hemisphere. An ESO target of opportunity program was activated to obtain FEROS spectra periodically through the early decline months. It was generally not possible to get spectra at regular intervals because of factors such as weather and commitment of the telescopes to large blocks of observing time for dedicated programs. Still, we were successful in acquiring series of spectra for many novae that reveal all the significant changes in line strengths, profiles, and continuum from the initial P Cygni-type spectrum at maximum light to the later forbidden emission line stage of the diffuse ejecta. Table 1 lists the novae for which we have FEROS spectra together with various characteristics of the novae, and the date and time from maximum visible brightness when each of the spectra were acquired.

The observing program was very straightforward. After a discovery announcement we incorporated every observable Galactic nova into our program. We estimated exposure times based on the reported visible brightness of the novae, normally taking three exposures that ranged from 60-1800 sec each that allowed us to median filter the exposures in the data reduction process. Standard stars were observed to determine the instrument response function (IRF) for flux calibration. The FEROS IRF is stable and on the few occasions when conditions were not photometric the compromised standard star calibration was augmented by use of mean extinction coefficients and the most recent IRF that had been determined in photometric conditions. Thus, absolute flux calibration may be in error for some of the spectra, but the relative fluxes should be correct. Wavelength calibration was done using standard lamps and procedures, resulting in a wavelength scale that is extremely stable and accurate to within 0.02 Å, based on the measured wavelengths of interstellar absorption features in the different orders.

The spectral evolution of novae has been well documented and is understood in terms of the evolving conditions in the expanding ejecta. Initially a photosphere forms in the optically thick gas, which becomes less dense and optically thin as it flows outward. The initial continuum and P Cygni line profiles from the expanding gas gradually decline and evolve to a weaker continuum with increasingly prominent emission lines. The properties of the broad absorption features observed near maximum light were studied long ago by Payne-Gaposchkin (1957) and McLaughlin (1960), who documented the evolution of broad H I Balmer, Fe II, and Na I profiles that consisted of "multiple absorption systems whose [expansion] velocities tend to increase with time" and that often "break up into smaller velocity components."

Changing spectral characteristics are very evident in our high resolution spectra, and we show several outstanding examples of the time evolution of the absorption associated with the Na I D lines in Figure 1. Near maximum light many novae show several Na I D absorption systems. The features appear with negative radial velocities, usually larger than 400 km s$^{-1}$ from the binary velocity, as defined by the subsequent forbidden emission lines, and they usually evolve blueward with time. The absorption features tend to broaden and sometime break up into discrete kinematical components as they are accelerated outward, gradually weakening in time. New Na D-doublet absorption features occasionally appear with lower expansion velocities. Corresponding absorption in the Balmer and Fe II lines shows similar behavior although less pronounced than the D lines because they are dominated by stronger emission components.

The outward acceleration of absorption systems is believed to be driven by a postoutburst wind or radiation from the white dwarf (Kovetz 1998; Hachisu & Kato 2005). As soon as forbidden emission lines emerge from the expanding ejecta the absorption disappears, likely due to both dissipation of the absorbing gas and increasing ionization by the WD wind and radiation. This transition usually occurs over a timescale of less than two weeks and in high resolution spectra is signaled by an evolution of the broad Na I 'D' P Cygni profile to the He I λ5876 emission line. Prior to this transition the bulk of the strong emission feature is to the red of the strong interstellar Na I D absorption, whereas afterward it is to the blue. This morphing of the Na I D → He I λ5876 emission signals a fundamental change in the nova spectrum from a P Cygni absorption spectrum to an emission-line spectrum, and it represents the disappearance of the nova photosphere.

One of the remarkable features of our FEROS spectra is that more than two-thirds of the novae show dozens of well-defined absorption lines near maximum light that have no emission counterpart and that punctuate the spectrum in the early decline period before forbidden emission lines have appeared. Most of the absorption lines are concentrated in the wavelength region 4000-6000 Å and can be identified as low excitation heavy element transitions. The more prominent systems have relatively narrow lines, e.g., 40 km s$^{-1}$, compared with the strong, broad P Cygni profiles that have widths greater than $10^3$ km s$^{-1}$, and they therefore are not detected at moderate spectral resolution. We are not aware of these systems having been studied before. McLaughlin (1960, p. 608) noted the occasional presence of Ti II and Cr II lines in some novae, but he did not mention any analysis or interpretation of these features. They possess a wide range of line widths, and although most of the novae do show the systems for a period of weeks after maximum light before they weaken and disappear, they are prominent in only a few novae which have a system with a small velocity dispersion, i.e., narrow, pronounced lines. Most, and perhaps all, of the transient absorption systems have as one of their strongest lines one of the Na I D absorption doublets that appear in novae at maximum light. The lines are not very deep, typically having residual intensities at line center greater than 70%, and it is the nature and origin of these absorption lines that we focus upon here.

## 3. Transient Heavy Element Absorption Systems

### 3.1. Observational Characteristics

During the first weeks after outburst the majority of novae show a continuum with broad P Cygni profiles of H I, Fe II, Na I, and O I, that is produced by an expanding photosphere of optically thick ejecta from the outburst. Expansion velocities are typically of order 1-2×$10^3$ km s$^{-1}$, and the low ionization of the spectral features is due to adiabatic cooling of the ejecta as they expand (Arnett 1979). In the days following the outburst the great majority of novae show strong Na I λλ5890,5896 'D-doublet' absorption that often has multiple absorption components representing different radial velocities. These normally consist of a strong absorption feature near the blue edge of the broad emission component, having $v_{exp}>1,500$ km s$^{-1}$, and which is also present in the Balmer and Fe II lines. This absorption is the P Cygni component associated with the broad emission from the ejecta.

Additional Na I D absorption doublets are narrower and have radial velocities corresponding to smaller expansion velocities of 400-1000 km s$^{-1}$. Such systems, several of which are evident in Fig. 1, are unambiguously identified from the D-line doublet spacing and relative strengths. Their absorption is usually superposed on the broad Na D emission component. These narrower D-line absorption components of lower expansion velocity originate from the same gas that produces the numerous absorption lines.

Our FEROS data reveal a large number of discrete absorption lines that permeate the spectra in the weeks following outburst and that are not associated with any transitions previously identified in novae. Figure 2 presents a montage of spectra of ten novae from our survey that show these absorption features, and a diversity of line strengths and widths is evident among the different novae. The spectra are presented roughly in order of decreasing line widths. The radial (expansion) velocities of the systems, relative to the nova binary velocity defined by subsequent forbidden line emission, fall within the range 400-1000 km s$^{-1}$. The FWHM line widths show a wide range of values between 35-350 km s$^{-1}$, reflecting large variations in internal velocity dispersion of the absorbing gas from nova to nova. The strengths of the absorption lines are such that equivalent widths range from the continuum noise level at the weak end to line center depths that sometime exceed 60% of the continuum intensity for the stronger lines. For a few of the novae in our survey we were successful in obtaining sequences of spectra that show the evolution of the absorption lines, and these are presented in Figure 3. The lines tend to broaden and move blueward with time before disappearing.

We have systematically identified the lines in the richest absorption systems, which are those belonging to Nova LMC 2005 and V378 Ser/05, both of which have numbers of clearly detected lines that far exceed any of the other systems we have observed. The radial velocities derived for the Na I D absorption lines in each nova were used to correct the observed absorption wavelengths to rest values, and identifications were sought for each radial velocity that resulted in a consistent set of transitions expected from a diffuse gas. We found that reasonable identifications for almost all lines generally resulted from one of the Na D radial velocities, usually that corresponding to the sharper Na D absorption system. The large majority of the absorption lines can be identified as low excitation Sc II, Ti II, V II, Cr II, & Fe II transitions. We refer to these absorption line systems which have observable lifetimes of order 2-8 weeks as *transient heavy element absorption* (*THEA*) systems.

The novae LMC 2005 and V378 Ser/05 possess the best defined THEA systems partly due to the high S/N of their spectra, but also because the absorbing gas in these two novae has the smallest velocity dispersions of all our objects, resulting in the narrowest lines among the novae sample. Nova LMC 2005 has by far the largest number of absorption lines of all the systems we have detected. The great majority of lines that we have identified in the other novae are present in the Nova LMC 2005 spectra. The absorption lines identified in LMC 2005 may therefore serve as a master list of the transitions we observe in THEA systems. A detailed spectrum of Nova LMC 2005 is shown in Figures 4 and 5 together with line identifications over the wavelength region 4000-5500 Å. Although the relative line strengths vary between different novae, and with time in each nova, this system, whose radial (expansion) velocity with respect to the nova system is –435 km s$^{-1}$, may be taken to be representative of THEA systems in novae. Note that a few lines have not been identified. Whether they belong to a system of another radial velocity or to an element with low solar abundance which happens to be enhanced, or simply have not been observed yet in the lab is not clear. In at least one nova, V2574 Oph/04, there are several strong Na D absorption systems after $+3^d$ (see Fig. 1b) which belong to separate THEA systems that differ in radial velocity from each other by 415 km s$^{-1}$. These two absorption systems are similar to each other and consist of many of the same Sc, Ti, and Fe transitions.

All of the identified THEA lines originate from levels of low excitation potential, with $\chi_{exc}$<4 eV, in singly ionized species. The lines are those that arise from a solar composition gas of low excitation conditions, i.e., the same transitions that occur in the spectra of late-type stars.

The statistics of the THEA systems are as follows: of the 15 novae observed near maximum light in our survey and listed in Table 1, 12 of them were observed to have transient heavy element absorption lines for at least one epoch. We have marked the dates of each of the spectra in Table 1 with an asterisk for those epochs where heavy element absorption lines were observed. Because of different S/N and sampling epochs of our spectra the statistics are consistent with every nova having a THEA system at some epoch. Interestingly, the two novae in our sample for which heavy element absorption lines were not detected in spite of our good temporal coverage with high S/N spectra, V382 Vel/99 and V5115 Sgr/05, are the two novae showing the most rapid declines in brightness. It may be that by the time of our first spectra of these objects at +5$^d$ and +6$^d$, respectively, the THEA gas had already dissipated.

### 3.2. Excitation Temperature

The relative strengths of the absorption lines originating from levels of different excitation potential of the same ion can be used to derive the excitation temperature of the gas at various epochs as it is accelerated. It is of interest to determine if heavy element absorption disappears because $T_{exc}$ drops below the heavy elements condensation temperature $T_{cond} \sim 1500$ K due to depletion onto dust. The best lines to use for this exercise are those that have good S/N, appear to be unblended, which are not too strong, i.e., possibly saturated, and which originate on levels of widely differing excitation potential. We focused attention on lines for which other members of the same multiplet were also observed so we could verify that the relative equivalent widths were those expected from the log (gf) values as a validation of the f-values and the equivalent width measurements. A list of the transitions we have used in our analysis of temperature and abundances, together with their measured equivalent widths, line widths, and f-values is given in Table 2.

For two optically thin transitions originating from levels i and j of the same ion, it is straightforward to show that the excitation temperature can be written

$$T_{exc} = \chi_{ij} \left\{ k \, \ln \left[ (W_\lambda^i g_j f_j \lambda_j^2)/(W_\lambda^j g_i f_i \lambda_i^2) \right] \right\}^{-1}, \tag{1}$$

where $W_\lambda^j$ is the equivalent width of the absorption line from level j>i, $\chi_{ij}$ is the difference in excitation potential of levels j and i, and $g_j$, $f_j$, and $\lambda_j$ are the statistical weight, f-value, and wavelength of the transition from level j upward.

Although the majority of novae in our survey display a THEA system in at least one epoch of observation, the best system for which we have good data with high S/N and unblended lines to determine the excitation temperature and relative abundances of the absorbing gas with some confidence is that of nova LMC 2005. Its absorption system shows little change over the interval of six weeks during which it was detected (see Fig. 3), and we selected the 5 Dec 2005 epoch for detailed analysis. The same-ion line pairs that best satisfy the requirements for analysis in this nova are (a) Ti II λλ4012.37 ($\chi_{exc}$=0.6 eV) & 4163.64 ($\chi_{exc}$=2.6 eV), (b) Sc II λλ4415.56 ($\chi_{exc}$=0.6 eV) & 5526.81 ($\chi_{exc}$=1.8 eV), and (c) Fe II λλ4491.40 ($\chi_{exc}$=2.8 eV) & 6416.92 ($\chi_{exc}$=3.9 eV). Using the measured equivalent widths given in Table 2 with f-values from Kurucz' (2008) compilation we derive values for the excitation temperature of $T_{exc}$=8573 (Sc II), 9766 (Ti II), & 11822 (Fe II) K for the primary THEA system of LMC 2005. Not surprisingly for systems in which absorption is observed from levels 3-4 eV above the ground state an excitation temperature of $10^4$ K is found, characteristic of the kinetic temperatures of H II regions. This temperature for the outer diffuse circumbinary gas is higher than that normally associated either with dust formation or with the expanding photosphere of novae ejecta, where $T_{phot}$<5,000 K (Warner 1995; Hauschildt et al. 1997). The relatively warm temperature may be a relic of photoionization from the hot white dwarf in the time immediately prior to the outburst.

### 3.3. Element Abundances

Our spectral resolution of 48,000 (6 km s$^{-1}$) easily resolves the absorption lines in the heavy element systems, which all have line widths of FWHM>35 km s$^{-1}$ at every epoch of the systems we have observed. The minimum residual intensity at line center for most of the lines we detect in the THEA systems exceeds 50% of the continuum intensity as can be seen in Figs. 4 & 5, since it is the strongest system we observed. Thus, the lines are not saturated unless they consist of numerous contiguous unresolved components, each of which has an intrinsic width less than ~3 km s$^{-1}$. This is sufficiently improbable that we can have some confidence that the absorbing gas is optically thin in the lines. The equivalent width of an optically thin absorption line in diffuse gas is (Spitzer 1978)

$$W_\lambda = \pi e^2/(m_e c^2) \ N_i \ \lambda^2 \ f_{ij} ,$$ (2)

where $N_i$ is the column density of absorbers in the lower level i of the transition. We have taken the measured equivalent widths and oscillator strengths of lines in Table 2 together with an assumed temperature of $T_{exc}=10^4$ K to calculate the column densities of the parent ions of the THEA lines in LMC 2005. The absorption originates from levels having a range of different excitation potentials, therefore we assume a Boltzmann distribution for the column densities of the levels to compute the total ion column density, $N_{ion}$, for each ion, and this quantity is what is shown in the final column of Table 2. The average of values determined for each ion have then been taken to determine the relative abundances of the heavy elements, assuming the fraction of singly ionized species to be the same for all elements, and these results are given in the bottom row of Table 3. The abundances are shown relative to Fe, which has been arbitrarily normalized to the solar value with respect to H. Because H absorption is highly saturated in our spectra, the Fe/H abundance cannot be determined from our spectra.

The column densities for the transient system in LMC 2005 can be used to calculate a rough mass estimate for the absorbing gas. If one assumes (1) a radius of 30 AU for the system, (2) a covering factor of ~30% for the gas, since it is seen in most novae, (3) a solar Fe/H ratio, and (4) a column density of $10^{18}$ cm$^{-2}$ for the Fe II in a typical nova, the mass of the THEA system is of order $10^{-5}$ M$_\odot$.. This is at the low end of the mass required in nova calculations to initiate the thermonuclear outburst on a WD (Starrfield et al. 2005; Yaron et al. 2005), although most of the mass may be ejected from the system and does not accrete onto the WD. This estimate is very approximate, of course, but it does indicate that mass ejection episodes by the secondary star could be the dominant form of mass transfer even over long periods of time in cataclysmic variable systems.

There are uncertainties in the abundances due to assumptions in the analysis and it is hard to quantify all of them. They involve (a) an assumed mean temperature of $T_{exc}=10^4$ K for all the ions, (b) f-values for a number of transitions which are difficult to determine experimentally and therefore have been computed from theory, (c) assuming the fraction of singly ionized species to be the same for all the heavy elements, and (d) no selective gas depletion among any of the elements due to condensation into dust. With regard to the latter two points, not only is the gas temperature much higher than $T_{cond}$, but the condensation temperatures of all of the heavy elements we have observed are similar, and should not lead to differences in relative depletions. Also, the ionization potentials of the first three stages of the elements we observe are all quite similar, so the relative abundances of the singly ionized species should not be very different from those of the elements.

Taken together we estimate that the error in the calculated abundance for any one of the THEA ions should be less than an order of magnitude. There is a clear pattern among the group of elements having atomic numbers less than Fe, viz., Sc, Ti, V, & Cr, which all have abundances relative to Fe that are appreciably above solar values. This systematic result gives some credibility to the idea that (Sc,Ti,V,Cr)/Fe abundances may be enhanced above solar values in the preoutburst circumbinary gas. We are not aware of any nucleosynthesis scenario that explains this general abundance pattern we find for the THEA systems. Because the errors in our analysis may be large the important question of abundances must be addressed by further study of more of these systems in future novae.

### 3.4. Interpretation

The apparent lower limit to the expansion velocities of the absorption systems of ~400 km s$^{-1}$ may be related to the escape velocity of the secondary star in the sense that gas ejected below a certain velocity is likely to fall back into the binary system. The high velocity limit to the accelerated systems is dictated by the WD postoutburst wind or radiation field. The gradual weakening of the absorption may be due to dissipation of gas by the accelerating wind or to an increase in ionization caused by interaction with the accelerating wind and radiation.

Are the metal systems produced in the outburst, or have they been produced prior to it and originate in gas whose accretion may have been the cause of the outburst? Given that at maximum light most novae show evidence of Na I D absorption systems having expansion velocities in the range 400-1000 km s$^{-1}$, it is possible that the outburst has produced these systems in the days between the thermonuclear runaway and maximum visible luminosity. However, inasmuch as the prominent Balmer, Fe II, and Na I D P Cygni features at maximum light originate in the more rapidly expanding outburst ejecta, the fact that the lower velocity heavy element absorption lines are superposed upon the P Cygni emission features, as is clearly seen for the H$\gamma$ emission component in most of the spectra in Fig. 2, requires that the THEA systems originate outside of the photosphere. Thus, the heavy element systems, which have lower expansion velocities than the outburst ejecta, must pre-exist the outburst.

Since most THEA systems experience outward acceleration, yet they are located outside the outburst ejecta, it raises the question as to what the accelerating mechanism is. Since the ejecta are located inside the transient systems, the WD wind would not be expected to have reached the outer THEA gas. A possible acceleration mechanism might be $\gamma$-rays emitted by proton-capture reactions associated with the outburst. Or, the ramp up to the outburst may have produced a WD wind resulting from the very high surface temperature, and this wind interacts with the THEA gas. A mechanism must also be invoked to account for the acceleration of the outburst ejecta, since the P Cygni absorption components of the strong Balmer and Fe II lines also show a steady migration to bluer wavelengths, as can be seen for the broad H$\gamma$ P Cyg absorption components in Fig. 3. Whatever the mechanism is, it is clear that a substantial fraction of the outer heavy element absorption systems experience acceleration after the outburst

Several facts point to the secondary star as the origin of the THEA gas: (1) expansion velocities of ~400-1000 km s$^{-1}$ are more characteristic of the secondary star escape velocity than the much higher WD escape velocity, and (2) it seems improbable that the heavy elements, which do not participate in the nuclear reactions of the nova outburst, are from the WD because they are expected to selectively diffuse into the WD interior. That said, one cannot absolutely rule out the WD as the origin of the THEA gas. Indeed, even if the source of the heavy element gas is mass loss by the secondary star, that material could have been transferred to it by the WD during an earlier common envelope stage. On balance, we interpret the high fraction of novae that exhibit expanding THEA gas at maximum light to be indicative of circumbinary gas ejected by the secondary star prior to the outburst, possibly episodically.

The fact that multiple Na I D line absorption systems are common suggests that there may be episodes of mass loss in the secondary star, presumably due to structural changes associated with its evolution in the close binary system, e.g., pulsations or instabilities. A fraction of the ejected gas is likely not to escape the binary system, but rather fall back onto the stars from dissipation of angular momentum in turbulent interactions, and the accreted gas may well be sufficient to trigger the outburst on the WD. There has long been a debate as to the cause of secondary outbursts only weeks after the initial nova outburst. Mass ejection episodes from the secondary star could initiate this activity. Whatever the cause, secondary maxima do require an explanation for the rather short interval of weeks between the initial and secondary outbursts that follow the much longer interval of years of quiescence that has preceded the initial primary outburst. With the outer layers of the WD still hot from the initial outburst, a second episode (or residual accretion from the first episode) of mass ejection from the secondary star provides additional pressure, heating, and fresh H to fuel another p-capture episode in the WD outer layers that produces renewed energy generation in the degenerate gas in a matter of days rather than centuries.

We present a schematic diagram showing the possible geometry of the different ejecta components of novae at outburst in Figure 6 that is consistent with the observations reported here. A mass ejection episode by the secondary star in a short-period binary system results in some fraction of gas being lost from the system (THEA gas, in red), primarily in the binary plane but with eventual diffusion out of the plane. Some gas does not escape the system (in red) and accretes onto the WD, triggering the nova outburst. The outburst results in the high velocity ejection of the outer layers of the WD (in black) that gives rise to the photosphere with its strong continuum and P Cygni profiles, and which overtakes the outer, slower moving THEA gas. For an assumed lifetime of the THEA gas of roughly one month and a velocity difference between the THEA and ejecta gas of $10^3$ km s$^{-1}$, the collision between the two gaseous components, which terminates the transient absorption system, takes place at a distance of order 10-100 AU from the two stars. We note that the disappearance of the heavy element absorption lines occurs around the time that the early 'permitted' P Cygni spectral phase changes to the 'auroral' forbidden emission phase (Williams et al. 1991), so this transition may be facilitated by the interaction between the two gaseous systems.

The acceleration of THEA systems can be explained by either wind or radiation from the hot white dwarf as long as the covering factor of the primary WD ejecta is small, as this allows the acceleration mechanism to pass through the ejecta and reach the more distant THEA gas. There is every indication that novae ejecta are very inhomogeneous. Old resolved nova shells, such as T Pyx (Shara et al. 1997) and GK Per/1901 show extremely inhomogeneous, clumpy structure. The Na D absorption lines in novae are always observed to be saturated inasmuch as the doublet ratio of equivalent widths is closer to 1:1 than the 2:1 ratio of the f-values. Yet, the central depths of the D-doublet never approach zero intensity as is expected for a saturated line. This requires the D-line absorbing gas to incompletely cover the continuum radiation source. The same is true of the THEA gas: some of the strengths of Ti II and Sc II lines do not correspond to the ratio of their (admittedly uncertain) f-values, indicating some saturation. Yet, the residual intensity of the lines exceeds 70%. This suggests that the THEA gas has a covering factor of less than 0.5. Thus, both the WD ejecta and THEA gas must be very clumpy.

The existence of two separate interacting gas systems might explain the large variations in the characteristics of dust in novae, some of which show virtually no observable evidence for dust formation whereas others show strong optical absorption and an appreciable fraction of the postoutburst luminosity radiated in the IR by dust (Gehrz 1988; Bode & Evans 1989). The ejecta are cooler than the outer THEA gas due to their rapid adiabatic expansion. The nova ejecta are also likely to be more dense, so dust formation is most likely to occur in the ejecta, e.g., in pockets where the gas temperature drops below the condensation temperature. Dust will form where proper conditions occur, but it would likely be destroyed by the heating that results when the two gas systems collide. Thus, the determining factor of the extent to which dust does or does not form in a nova may depend on whether the conditions for dust formation in the WD ejecta occur before it collides with the outer layer of THEA gas.

## 4. Type Ia Supernovae

It is widely believed that SNe Ia, like classical novae, may occur in close mass transfer binaries, however this has proven difficult to establish conclusively. Eclipsing systems having periods less than one day have been observed in some old novae with high orbital inclinations, but have not been observed in quiescent SNe Ia systems. The lack of such observations for SNe is understandable due to the difficulty of making such a detection in extremely faint extragalactic sources in crowded fields. If SNe Ia are indeed cataclysmic variable systems it is possible that the supernova outburst is triggered by a discrete episode of mass transfer from the secondary star, and that the same heavy element absorption lines observed in novae at the time of maximum might also be present in the spectra of SNe Ia. Patat et al. (2007) have found spectroscopic evidence that they interpret in terms of circumstellar material around SN Ia 2006X.

Because of the higher expansion velocities of SNe Ia compared with those of novae, the principle SNe ejecta are likely to collide with circumbinary gas that gives rise to the heavy element absorption lines sooner after the outburst than observed in novae, perhaps even before SNe maximum light. It would therefore be instructive to acquire high-resolution spectra of SNe Ia as soon after discovery as possible, and preferably before visual maximum. Detection of discrete heavy element absorption systems like those reported here for novae could provide strong evidence that the SNe Ia phenomenon is associated with mass transfer onto a white dwarf. Differences in the characteristics of novae vs. SNe Ia circumbinary absorption systems could reveal important parameters that determine which type of outburst results from mass transfer episodes. In particular, episodic mass transfer involving relatively large amounts of secondary star mass might take place in conditions that enable a WD to exceed the Chandrasekhar limit and collapse in spite of the steady loss of its mass that has been predicted to occur from models of repeated nova outbursts.

## 5. Summary

High resolution spectra of novae at the time of outburst reveal many absorption lines that are identified with Fe-peak and s-process elements. The lines originate from warm gas that is expanding outwardly from the novae at velocities of 400-1000 km s$^{-1}$, and they gradually weaken and disappear in roughly 2-8 weeks. The gas originates before the outburst, and almost certainly comes from the secondary star. Most of the heavy elements observed appear to be enhanced with respect to Fe.

If the secondary stars of CVs do experience episodic mass loss it is likely that they do so at times that do not lead to an outburst. Confirmation of this scenario could come from observing old nova systems with high resolution spectroscopy to see if THEA systems are detected in novae in quiescence, between outbursts. A large telescope would be necessary to provide the necessary S/N for detection, but a large sample of old novae exists and THEA systems should be observable against the optical continua of CV accretion disks at quiescence.

**Figure 1**

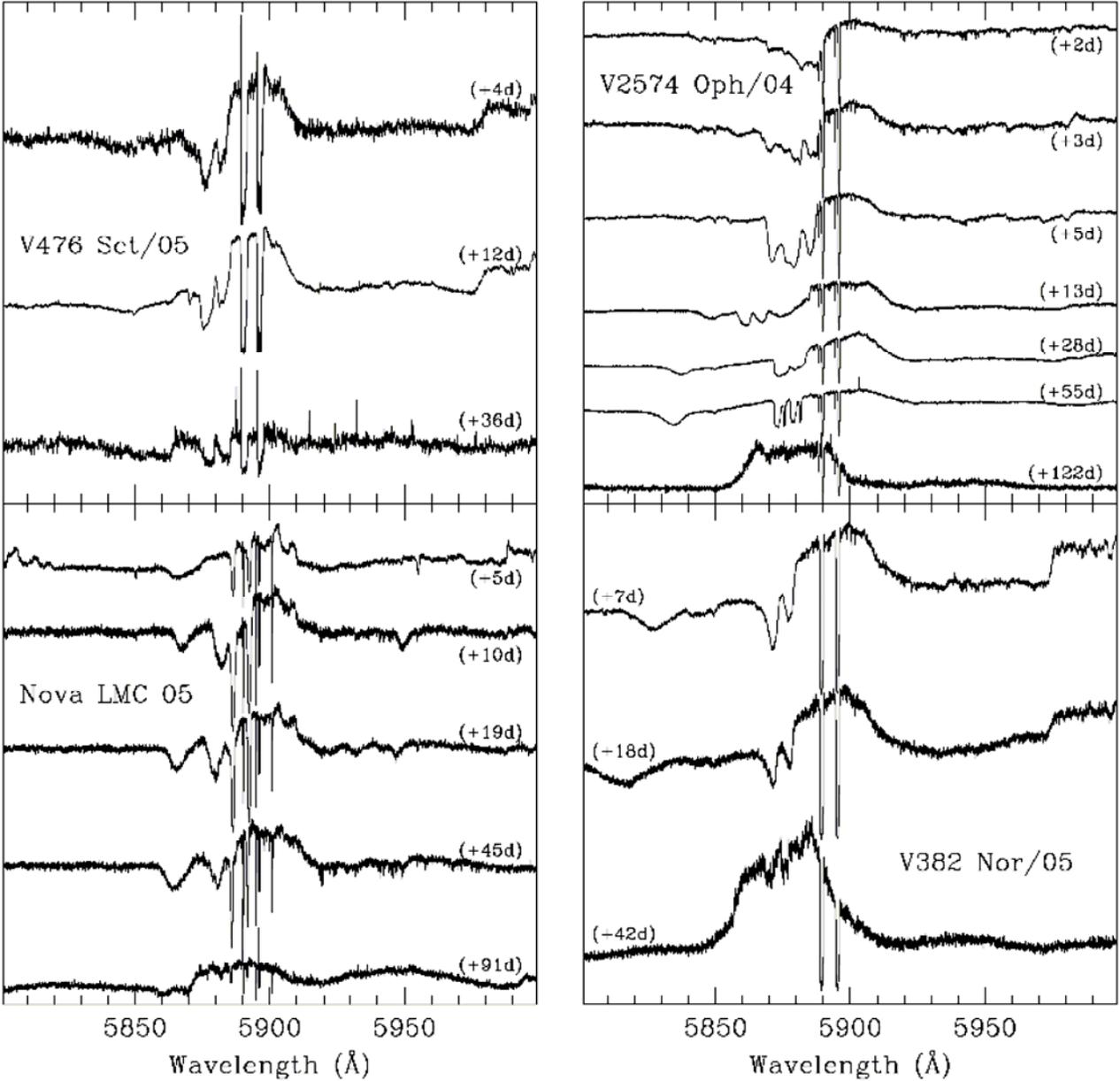

**Fig. 1**. The early evolution of Na I D absorption features in novae (a) V476 Sct/05, (b) V2574 Oph/04, (c) LMC 2005, and (d) V382 Nor/05. Note the blueward shift in radial velocity with time for some absorption components, and also the Na D absorption lines superposed upon the broad He I λ5876 emission line in V382 Nor, indicating that they originate outside of the emission-line region. The strong, sharp doublet absorption in each spectrum is due to the intervening ISM.

**Figure 2**

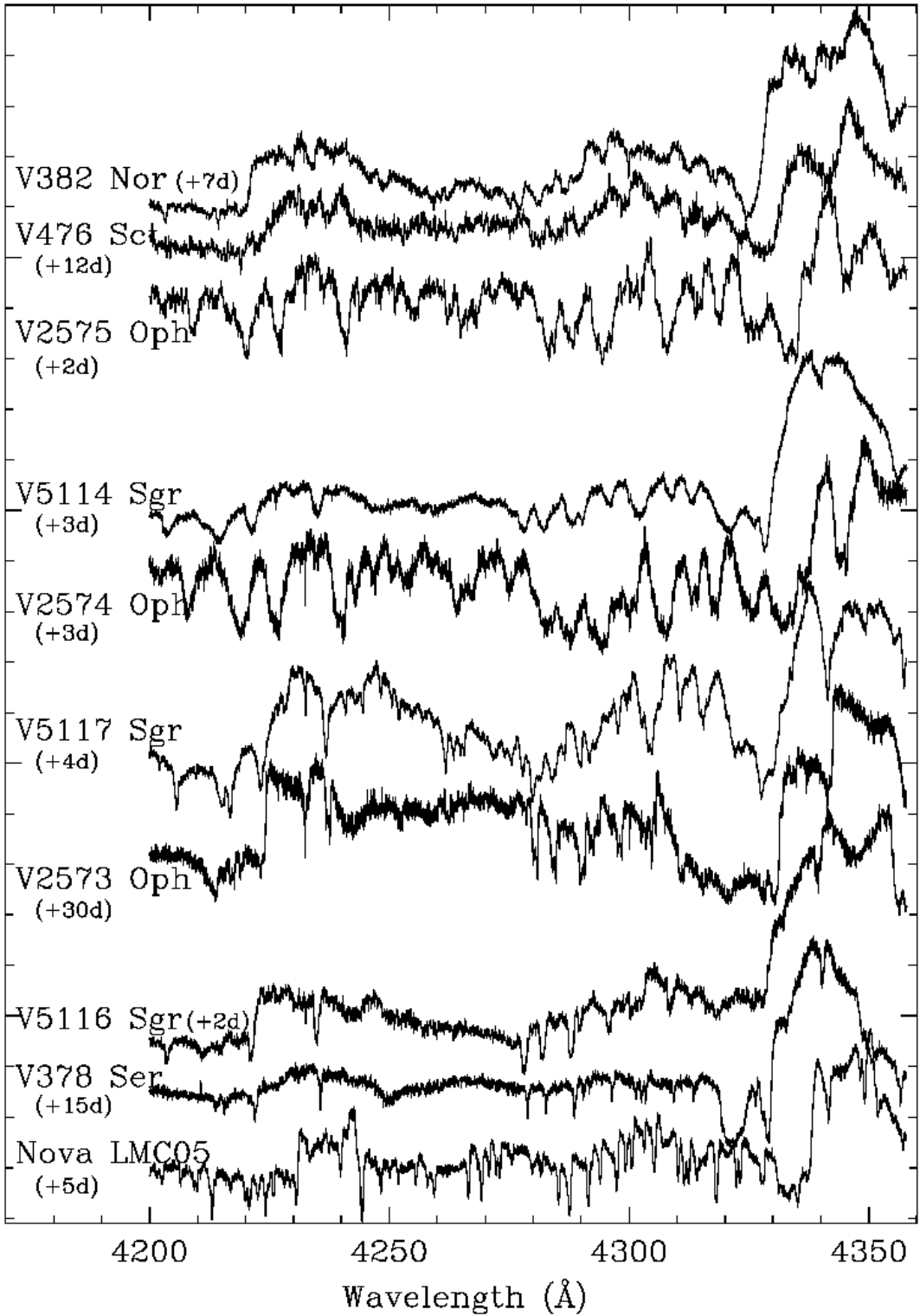

**Fig. 2.** FEROS spectra of ten of our survey novae near maximum light that show the numerous absorption lines in the blue spectral region that characterize high resolution spectra early in decline. The spectra are arranged in approximate order of decreasing line widths, from top to bottom. The absorption lines are identified with Sc II, Ti II, V II, Cr II, Fe II, & Sr II.

**Figure 3**

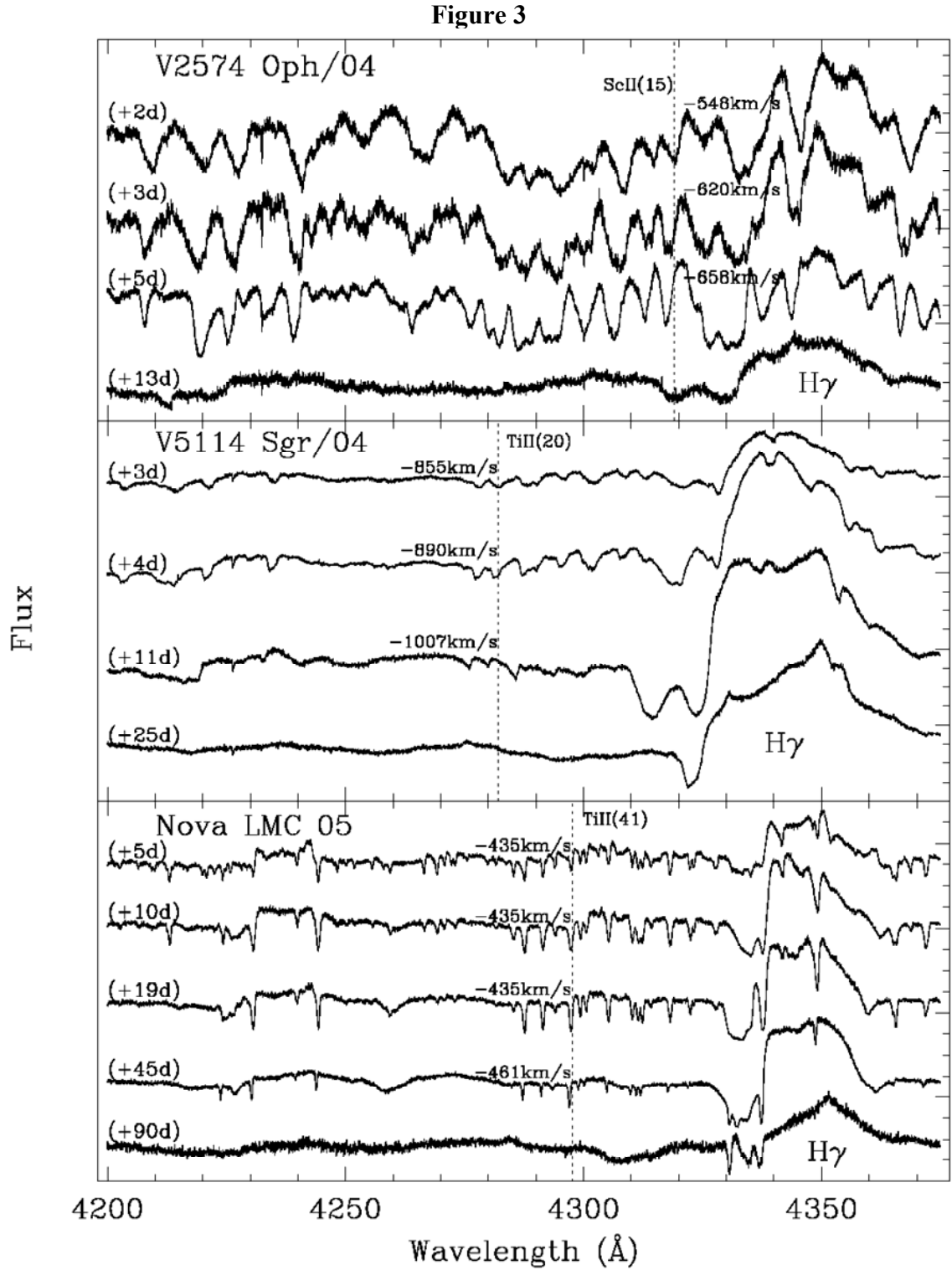

**Fig. 3.** Time evolution of the transient heavy element absorption systems in three well studied novae: (a) V2574 Oph/04, (b) V5114 Sgr/04, and (c) LMC 2005. The vertical dashed lines serve as a wavelength fiducial to highlight changing radial velocities of the systems, which are listed relative to the forbidden emission line systemic velocities of the novae. The V2574 Oph and V5114 Sgr absorbing gas is clearly accelerated outwardly with time, whereas the LMC 2005 absorption system shows little change in velocity over its 6-week lifetime.

**Figure 4**

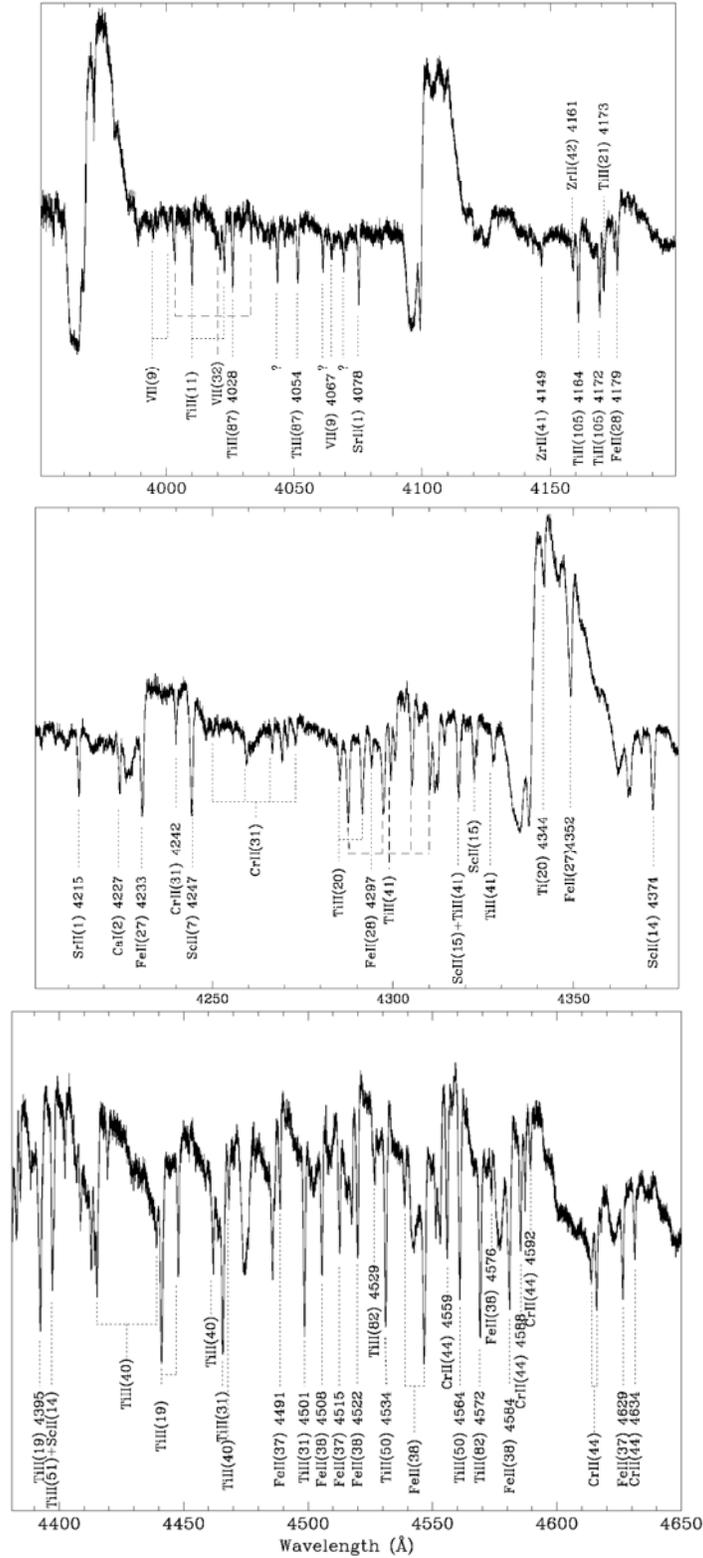

**Fig. 4.** The spectrum of nova LMC 2005 on 2005 Dec 1 between 3950-4650 Å, where many of the transient heavy element lines are concentrated. Line identifications are shown for this system, which is the most prominent transient absorption system we have observed in any nova.

**Figure 5**

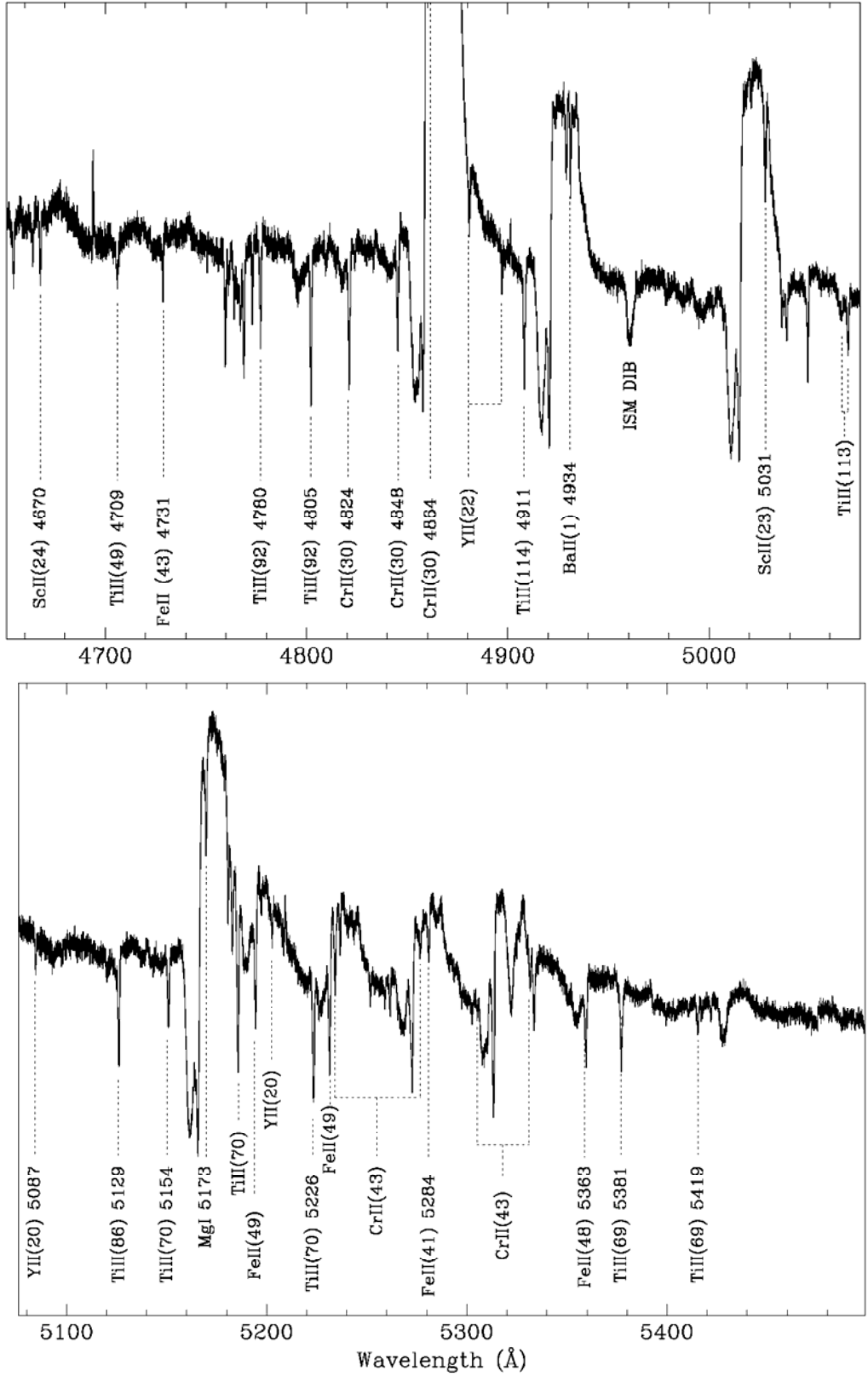

**Fig. 5**. Same as Fig. 4 except for the wavelength region 4650-5500 Å

.

**Figure 6**

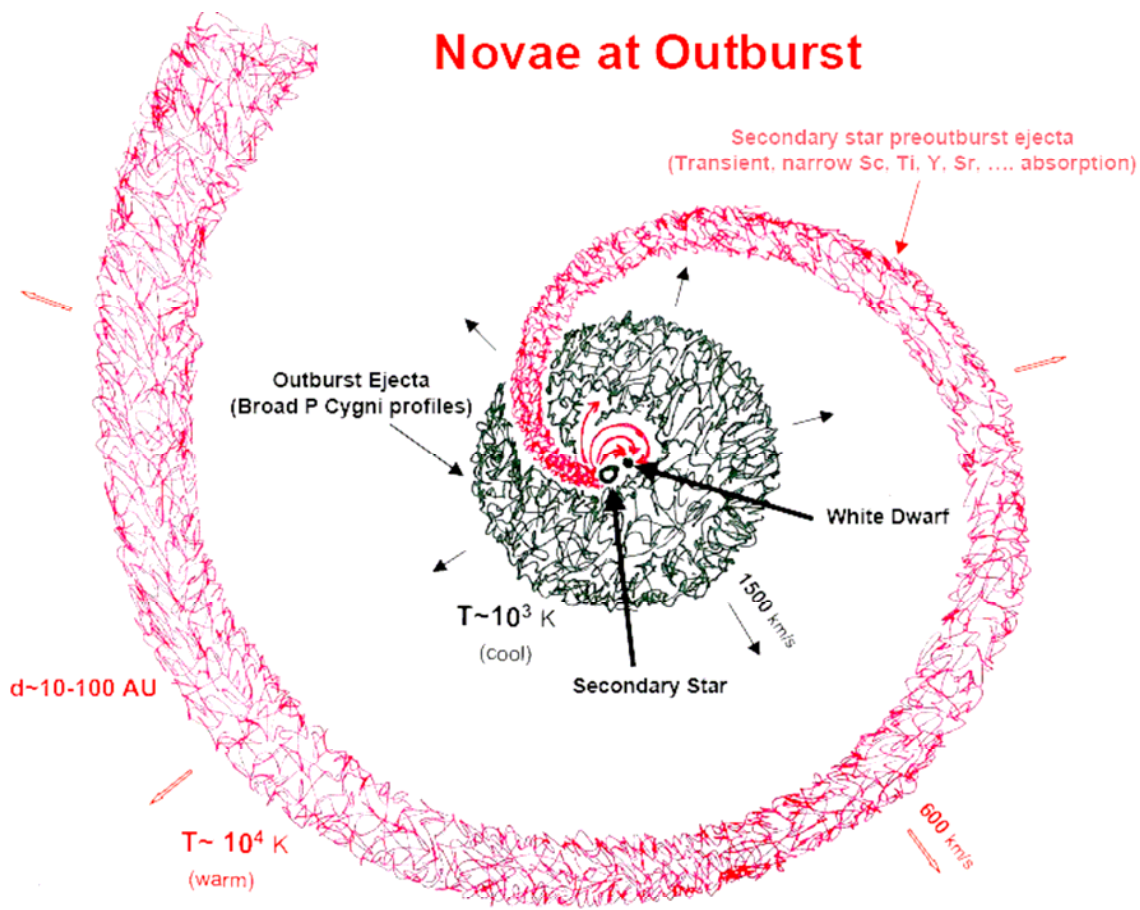

**Fig. 6.** A schematic representation of the gas associated with novae near the time of outburst. The red spiral represents material ejected by the secondary star before the outburst, some of which escapes and some of which is accreted onto the white dwarf. The central black sphere represents the ejecta of the white dwarf from the nova outburst. It produces the luminous photosphere and eventually collides with the outer gas from which the transient heavy element absorption systems originate.

## Table 1.  FEROS Spectra Epochs

| Nova | Date of Visible Maximum[a] | $t_3$[a] (days) | Dates of Spectra[a,b] (days after visible maximum) |
|---|---|---|---|
| V382 Vel/99 | 1999 May 23 | 10 | +5, +6, +7, +9, +33, +52, +57, +69, +184, +530 |
| V2573 Oph/03 | 2003 Jul 6 | 58 | +12*, +30*, +62 |
| V5114 Sgr/04 | 2004 Mar 15 | 22 | +3*, +4*, +11*, +25, +32, +57, +101, +190 |
| V2574 Oph/04 | 2004 Apr 15 | 35 | +2*, +3*, +5*, +13*, +28*, +54, +122, +128, +156 |
| V1186 Sco/04 #1 | 2004 Jul 5 | 34 | +35* |
| V1187 Sco/04 #2 | 2004 Aug 4 | 17 | +8, +11, +18, +33, +45, +56, +57 |
| V382 Nor/05 | 2005 Mar 18 | 18 | +7*, +18*, +42, +73, +113, +159 |
| V378 Ser/05 | 2005 Mar 21 | 85 | +15*,+39, +87, +88, +109, +157, +158, +188 |
| V5115 Sgr/05 #1 | 2005 Mar 30 | 13 | +6, +30, +47, +78, +79, +145, +147 |
| V1663 Aql/05 | 2005 Jun 10 | 25 | +7*, +28, +64, +77 |
| V5116 Sgr/05 #2 | 2005 Jul 5 | 23 | +2*, +21, +57, +80 |
| V476 Sct/05 #1 | 2005 Oct 1 | 26 | +4, +12*, +36 |
| LMC 2005 | 2005 Nov 26 | -- | +5*, +10*, +19*, +45*, +90 |
| V2575 Oph/06 | 2006 Feb 12 | 66 | +2*, 31, 33 |
| V5117 Sgr/06 | 2006 Feb 17 | 86 | +4*, +25* |

[a] Approximate since many novae are discovered after maximum light
[b] * = transient heavy element absorption lines present

**Table 2.  LMC 2005 THEA System Equivalent Widths and Column Densities**

| Line | log (gf) | $\chi_{exc}$ (cm$^{-1}$) | $W_\lambda$ (Å) | FWHM (km s$^{-1}$) | log $N_{ion}$ (cm$^{-2}$) |
|---|---|---|---|---|---|
| Zr II 3998.95 | -0.39 | 4506 | 0.052 | 52 | 14.15 |
| V II 4005.71 | -0.46 | 14656 | 0.253 | 63 | 15.52 |
| Ti II 4012.38 | -1.61 | 4629 | 0.351 | 53 | 16.21 |
| V II 4023.39 | -0.88 | 14556 | 0.112 | 49 | 15.57 |
| Ti II 4028.34 | -1.01 | 15257 | 0.301 | 49 | 16.21 |
| V II 4035.63 | -0.96 | 14462 | 0.101 | 46 | 15.61 |
| Ti II 4053.83 | -1.21 | 15266 | 0.302 | 54 | 16.41 |
| Sr II 4077.71 | 0.17 | 0 | 0.346 | 52 | 12.73 |
| Zr II 4149.22 | -0.03 | 6468 | 0.119 | 47 | 14.24 |
| Ti II 4163.65 | -0.4 | 20892 | 0.556 | 56 | 16.19 |
| Fe II 4178.86 | -2.48 | 20831 | 0.351 | 54 | 17.94 |
| V II 4202.36 | -1.75 | 13742 | 0.058 | 50 | 16.07 |
| Sr II 4215.52 | -0.14 | 0 | 0.297 | 45 | 12.95 |
| Fe II 4233.17 | -2 | 20831 | 0.604 | 62 | 17.68 |
| Cr II 4242.36 | -0.59 | 31219 | 0.145 | 47 | 15.72 |
| Cr II 4269.28 | -2.17 | 31083 | 0.087 | 41 | 17.07 |
| Ti II 4287.87 | -2.02 | 8710 | 0.204 | 55 | 16.58 |
| Sc II 4305.71 | -1.22 | 4803 | 0.051 | 41 | 14.53 |
| Sc II 4325.00 | -0.44 | 4803 | 0.254 | 56 | 14.45 |
| Fe II 4351.77 | -2.1 | 21812 | 0.543 | 91 | 17.77 |
| Sc II 4415.56 | -0.64 | 4803 | 0.217 | 51 | 14.56 |
| Fe II 4491.40 | -2.7 | 23031 | 0.213 | 55 | 18.02 |
| Ba II 4554.03 | 0.17 | 0 | 0.118 | 36 | 12.49 |
| Cr II 4558.65 | -0.66 | 31219 | 0.497 | 65 | 16.26 |
| Cr II 4618.80 | -1.11 | 32855 | 0.337 | 53 | 16.64 |
| Cr II 4634.07 | -1.24 | 32845 | 0.211 | 48 | 16.56 |
| Cr II 4824.13 | -1.22 | 31219 | 0.401 | 56 | 16.68 |
| Cr II 4848.24 | -1.14 | 31169 | 0.265 | 52 | 16.41 |
| Y II 4883.68 | 0.07 | 8743 | 0.132 | 58 | 13.68 |
| Y II 4900.12 | -0.09 | 8328 | 0.082 | 43 | 13.61 |
| Ba II 4934.08 | -0.15 | 0 | 0.104 | 38 | 12.68 |
| Y II 5087.42 | -0.17 | 8743 | 0.066 | 39 | 13.59 |
| Y II 5205.72 | -0.34 | 8328 | 0.053 | 43 | 13.62 |
| Ti II 5211.54 | -1.36 | 20892 | 0.067 | 47 | 16.03 |
| Sc II 5526.79 | 0.13 | 14261 | 0.408 | 54 | 14.46 |
| Ba II 6141.72 | -0.076 | 5675 | 0.225 | 56 | 12.29 |
| Fe II 6238.39 | -2.63 | 31364 | 0.149 | 59 | 18.03 |
| Fe II 6416.92 | -2.74 | 31388 | 0.157 | 59 | 18.14 |
| Ba II 6496.90 | -0.38 | 4874 | 0.154 | 46 | 13.15 |

### Table 3.  Temperature & Abundances of Nova LMC 2005 THEA System

| $T_{exc}$ (K) | 8,573 (Sc II)  9,766 (Ti II)  11,822 (Fe II) | | | | | | | | |
|---|---|---|---|---|---|---|---|---|---|
| **Relative  Abundances:** | | | | | | | | | |
| Element | **Sc** | **Ti** | **V** | **Cr** | **Fe** | **Sr** | **Y** | **Zr** | **Ba** |
| **LMC 2005 THEA Abundance** (log Fe ≡ 7.54) | 4.11 | 5.88 | 5.30 | 6.09 | 7.54 | 2.45 | 3.24 | 3.81 | 2.26 |
| **Solar Abundance** (Lodders  2003) | 3.15 | 5.00 | 4.07 | 5.72 | 7.54 | 2.99 | 2.28 | 2.67 | 2.25 |
| **(THEA System – Solar)** | 0.96 | 0.88 | 1.23 | 0.37 | --- | -0.54 | 0.96 | 1.14 | 0.01 |